\documentclass[aps,preprint,amsmath,amssymb]{revtex4}
\usepackage{graphicx} 

\begin{document}

\title{Tunable failure: control of rupture through rigidity}

\author{Michelle M. Driscoll$^{1}$}
\author{Bryan Gin--ge Chen$^{2}$}
\author{Thomas H. Beuman$^{2}$}
\author{Stephan Ulrich$^{2}$}
\author{Sidney R. Nagel$^{1}$}
\author{Vincenzo Vitelli$^{2}$}
\affiliation{ $^{1}$The James Franck Institute and Department of Physics, The University of Chicago }
\affiliation{$^{2}$ Instituut-Lorentz, Universiteit Leiden, Postbus 9506, 2300 RA Leiden, The Netherlands}


\begin{abstract}
We investigate how material rigidity acts as a key control parameter for the failure of solids under stress.  In both experiments and simulations, we demonstrate that material failure can be continuously tuned by varying the underlying rigidity of the material while holding the amount of disorder constant.  As the rigidity transition is approached, failure due to the application of uniaxial stress evolves from brittle cracking to system-spanning diffuse breaking.  This evolution in failure behavior can be parameterized by the width of the crack.  As a system becomes more and more floppy, this crack width increases until it saturates at the system size.  Thus, the spatial extent of the failure zone can be used as a direct probe for material rigidity.
\end{abstract}

\maketitle

Brittle solids typically fail suddenly and catastrophically under an applied load;  an initially microscopic crack will rapidly grow, and quickly traverse the entire sample, breaking it into two.  At the tip of this crack, the stress applied at the boundaries is highly focused into a small region, often termed the `failure process zone'.  It is in this small region that the rupturing and breaking of bonds actually occurs; in this zone, which generally depends on material toughness~\cite{Barenblatt,Eran}, dissipation and non-linearities dominate the dynamics~\cite{bonamy,alava}.  Outside this zone, linear elasticity characterizes the medium's response.  In very rigid solids, the process zone is microscopic and the highly non-linear dynamics within it are difficult to observe.  However, the size of the process zone increases vastly if sufficient disorder is present~\cite{ponson1,santucci,shek}. Thus disorder is one relevant parameter for describing the nature of solid failure under tension.

In this work, we demonstrate that mechanical rigidity is another equally important parameter that controls material failure.  The spatial extent of the process zone under loading is large for marginally rigid solids, e.g.\ materials which lie close to a rigidity transition\cite{alexander,thorpe,phillips,ohern,wyart,ARCMP}.  Examples of a marginally rigid solid include sparsely connected disordered elastic networks, loose granular packings and cellular materials constructed from slender beams.  In all these cases, the loss of rigidity is signaled by the vanishing of a linear elastic modulus.  

\begin{figure*}
\centering 
\begin{center} 
\includegraphics[width=0.95 \columnwidth]{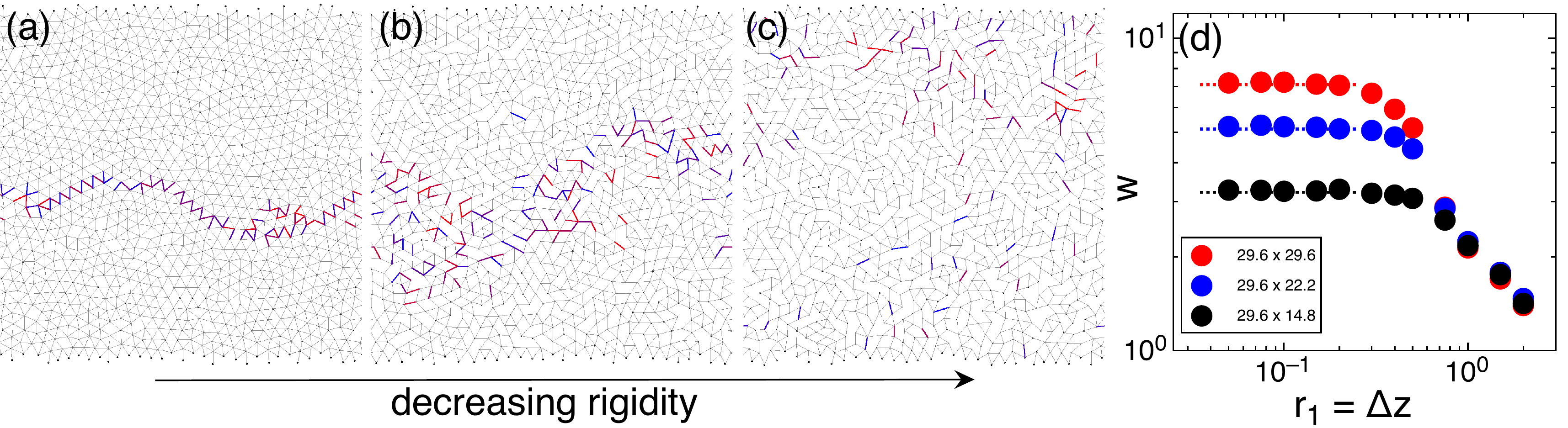} 
\caption{
\label{phenomena}
Transition in material failure due to changing material rigidity.  As bonds between nodes are removed from a random spring network, the failure behavior changes continuously from brittle cracking to diffuse breaking.  In the figures, the sample is stressed uniaxially in the vertical direction.  (a) A thin crack is observed in a rigid material.  Only a small amount of meandering occurs.  (b) The intrinsic width of the crack grows as the material becomes less rigid.  (c) The crack width saturates at the size of the system when the rigidity of the material decreases towards the limit where the bulk and shear moduli vanish.  (d) The width of the zone in which failure occurs, $w$, increases continuously as the rigidity, $r_1 \equiv \Delta z$, decreases and plateaus at a value $w_0$, indicated by the dotted line. Once the crack width reaches the system size, only diffuse breaking occurs.  The sample sizes are indicated in the legend: the horizontal width of the sample (perpendicular to the direction of applied stress) is kept constant at 29.6 and its vertical length (parallel to the pulling direction) varies from 14.8 to 29.6 (measured in units of the average interparticle spacing).  Simulations use dissipative Newtonian dynamics, as detailed in the  text. } 
\end{center}
\end{figure*}

Using rigidity as a control parameter, we can dramatically change how a material fails.  As illustrated in Fig.\ \ref{phenomena}a-c, it is evident that there are two distinct effects that give rise to a broadened failure zone.  First we note that, due to disorder, even an ordinary thin crack (Fig.\ \ref{phenomena}a) will meander horizontally in the direction transverse to the pulling direction as it traverses the sample.  This gives rise to a failure zone that, although thin at each point along its trajectory, fluctuates in the vertical direction parallel to the applied stress.  The second effect, which is the subject of the current paper, is that the intrinsic width of the failure zone grows as the rigidity of the sample is reduced as seen in Fig.\ \ref{phenomena}b.  Thus,  diffuse broken bonds vertically span a region around the center of the failure zone (perpendicular to the pulling direction). 
We define the intrinsic width as the distribution of breaks about the centerline of the failure zone.  As the intrinsic width of the failure zone grows, it dominates over the effect of meandering; the effect of meandering is no longer visible until the transverse width of the sample is much larger than the intrinsic width of the crack.  With decreasing rigidity, the crack width, and hence the process zone, continuously grows until it engulfs the entire sample as seen in Fig.\ \ref{phenomena}c.  This effect appears to be quite robust.  We find that the crack width can be tuned via rigidity in several distinctly different systems.  In all the cases we examine here, thin cracks in rigid systems turn into broad, diffusive failure as the initial, unstressed, system is prepared to be closer to a rigidity transition.

\section{Model systems and breaking protocols}

We study three distinct systems whose rigidity, measured by different geometric parameters, can be systematically controlled prior to the application of stress.  By comparing these three systems, we can isolate effects due to the material's geometry from those due to disorder.  Thus we can rule out that the observed behavior is solely due to changes in the degree of disorder. 

The first system, illustrated in Fig.\ \ref{phenomena}, is a simulation of a two-dimensional ($D=2$) disordered network of point masses connected by unstretched springs of equal stiffness whose average coordination $z$ can be varied.  Such a network is at the isostatic threshold when $z = z_c =2D=4$. At the isostatic point the number of constraints just balance the number of degrees of freedom  so that the network is poised on the brink of failure; if there are no redundant constraints, then the removal of just one bond creates a zero-frequency normal mode of vibration. \cite{maxwell1864calculation} Networks at non-zero $\Delta z \equiv z-z_c$ are prepared by randomly removing the bonds with highest coordination from an initial, highly-compressed, jammed packing of spheres until the target $z$ is attained (see \cite{wouter,ulrich} for details). This preparation procedure ensures that, upon lowering the geometric control parameter $r_1 \equiv \Delta z$, both $G$ and $B$ become vanishingly small, while their ratio, as well as the amount of disorder remains constant.     

Our second system is a simulation of a {\it perturbed} square lattice of point masses connected by equal-stiffness harmonic springs.  (The nodes of this square lattice are displaced from those of a perfect crystal by a small random vector  $0.1s$ ,where $s$ is the lattice spacing.)  Here the rigidity parameter is again $r_1 \equiv \Delta z$.  It counts additional springs (braces) added randomly along the diagonals \cite{mao}. The completely ordered square lattice without braces at $ \Delta z =0$ is isostatic and has a vanishing shear modulus, $G$, and a finite bulk modulus, $B$.  However, in our slightly disordered network, $B$ drops to zero, as does $G$.  Because each diagonal spring acts as an ``impurity'', the disorder in the local connectivity varies with $ \Delta z$.  

In both of these simulation systems, periodic boundary conditions are applied in the direction perpendicular to the uniaxial pulling direction.  Our numerical results are not sensitive to the choice of lateral boundary conditions for sufficiently large samples.  The samples are pulled quasi-statically and springs are removed whenever they are strained above a specified threshold value.  The quasi-static dynamics are obtained using two protocols: (i) in the random networks, we explicitly solve Newton's equations in the presence of microscopic dissipation as detailed in Ref.\ \cite{ulrich}; (ii) in the square system, we compute the linear response to an applied tension and remove the spring under the maximum stress before repeating. The process is stopped in (i) when the sample breaks into two parts, and in (ii) when the relevant linear elastic modulus drops to zero. 

Our third example is an experimental system consisting of a macroscopic, weakly disordered, honeycomb lattice which is pulled uniaxially. The lattice structure, laser-cut from plastic sheets, is characterized by the ratio of strut width, $x$, to strut length, $a$.  As in the braced square system, the nodes are displaced from those of a perfect lattice by random vectors with magnitude less than $0.1s$, where $s$ is the spacing of the underlying lattice.  The honeycomb lattice, with a coordination number $z= 3$, is over-constrained if both bond bending and central forces are present, but it is under-constrained if only central forces are considered.  One can tune the rigidity of this meta-material by lowering the aspect ratio of the struts, $x/a$. The bending energy and shear modulus $G$ are proportional to $\left(x/a \right)^3$, while the stretching energy and bulk modulus $B$ are proportional to $x/a$ \cite{book}. The parameter $r_2\equiv \sqrt{G/B}$ controls rigidity without altering structural disorder.  This definition of $r_2$ was chosen so that $r_2 \propto x/a$ for the honeycomb lattice.  As $r_2 \rightarrow 0$, the honeycomb lattice can increasingly accommodate compressions by beam bending rather than stretching.  For comparison, we have also studied triangular lattices with $z=6$ where $r_2$ does {\em not} measure rigidity.  The triangular lattice, as distinct from the honeycomb lattice, remains rigid as $x/a \rightarrow 0$ so that $G/B \sim $ constant as $x/a$ is varied.

In the experimental study, the side edges (perpendicular to the direction of pulling) are free.  The loading is applied using a strain-controlled translation stage (at a fixed displacement rate of 83 $\mu$m/s) while the spatial distribution of the broken bonds is recorded using a high-speed camera (at rates varying from 10 fps to 190,000 fps). The uniaxial pulling is continued until the samples have fully broken into two parts.

\section{Tunable Failure}

\begin{figure*}
\centering 
\begin{center} 
\includegraphics[width=0.95 \columnwidth]{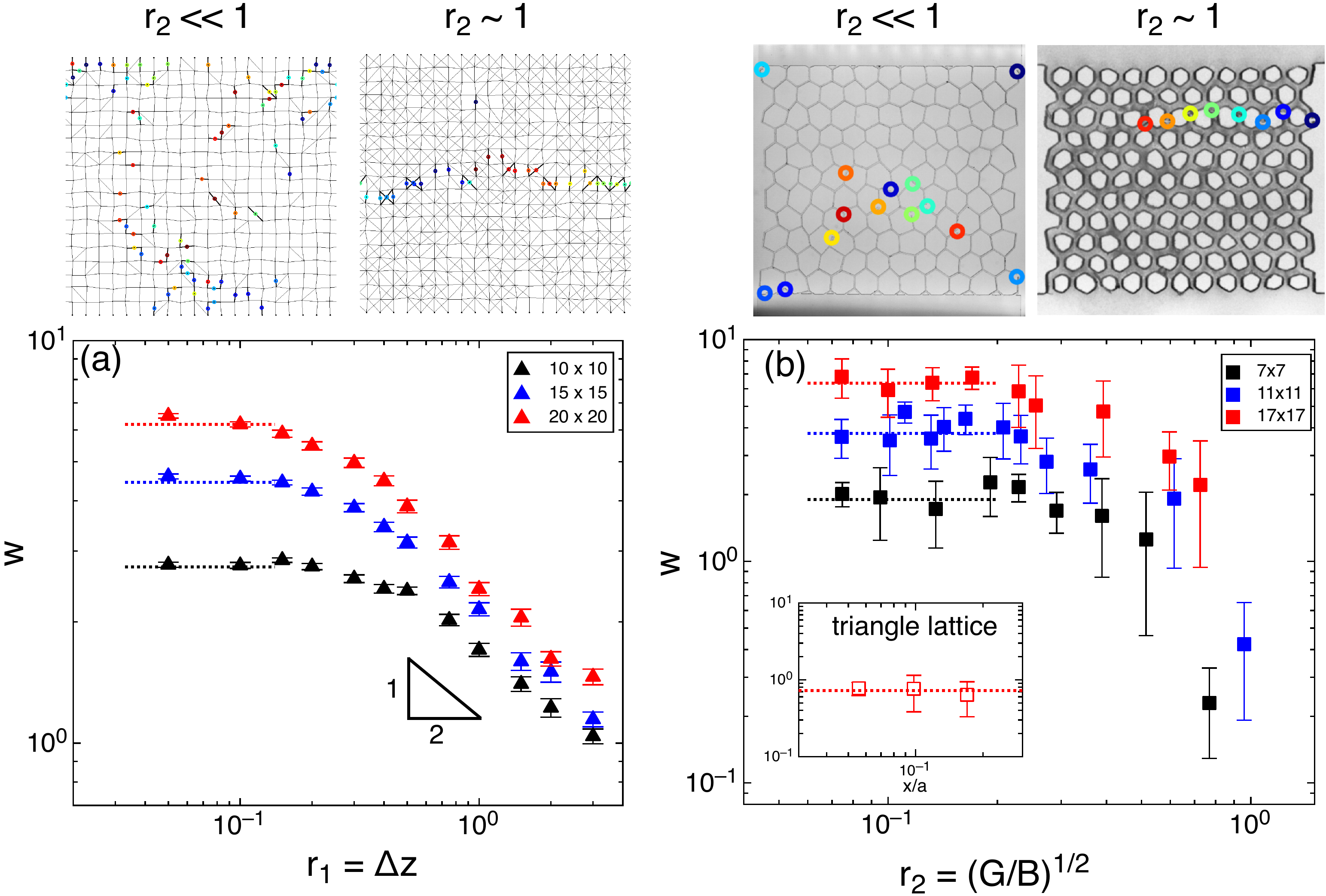} 
\caption{
\label{width}
Crack width versus rigidity for (a) simulations of perturbed square lattices with random braces and (b) experiments on perturbed honeycomb lattices.  The images above each graph illustrate the limiting behavior of the crack width, $w$, at small and large values of the rigidity parameter $r_i$.   As in Fig.\ \ref{phenomena}, as $r_i$ is decreased, the crack width, $w$, increases, but then plateaus at a value indicated by a dotted line. The inset in (b) shows $w$ vs $x/a$ for the triangular lattice, demonstrating that $w$ remains near 1 (in units of the lattice spacing) for all values of $x/a$. This indicates that the increase in crack width in the honeycomb lattice is related to the decrease in rigidity and not to the removal of excess material from the sheet or to the aspect ratio of the struts themselves.  The legends give the sample sizes in units of the length of the unit cell. } 
\end{center}
\end{figure*}

Material failure can be tuned by changing material rigidity.  In order to quantify the change from narrow crack to diffuse failure, we measure the dimensionless width of the damage zone, $w$, defined as the standard deviation of the vertical position of the broken bonds normalized by the lattice spacing.  Figures \ref{phenomena}d and Fig. \ref{width}a,b, show $w$ versus $r_i$ for various sample sizes, illustrating that failure is tunable in all three systems despite the obvious differences in their nature.  In all cases, a narrow crack is observed at high rigidity, similar to the failure behavior of a typical brittle elastic solid.  In this regime, bonds break in succession in a nearly straight line perpendicular to the direction of applied tension.  The small degree to which these thin cracks meander is governed by the disorder in the material \cite{bonamy}.  However, as the rigidity $r_i \rightarrow 0$, the crack width increases.  When $r_i \ll 1$, this width spans the entire system, creating a broad and diffuse failure zone.  In this state, bonds break in a nearly uncorrelated fashion until isolated damage zones eventually coalesce to  produce a percolating cluster of broken bonds across the sample.   

Even though there are large differences in the three systems (static vs.\ dynamic, central vs.\ bond-bending forces, nearly ordered vs.\ highly disordered), in all cases the failure behavior can be tuned continuously by varying the parameter $r_i$.  This remains true even though $r_i$ is controlled in very different ways in the experiments and simulations (changing strut aspect ratio vs.\ adding additional cross braces as defects or changing the connectivity of a highly disordered system).  Reducing $r_i$ marks the approach to an under-constrained mechanical state with vanishing elastic moduli.  This is accompanied by an increasing crack width.

In the case of the experimental system, as a check on the importance of the rigidity transition, we also used a triangular lattice (which remains rigid as $x/a \rightarrow 0$).  In this system, $w$ remains constant as $x/a \rightarrow 0$, as shown in the inset of Fig.\ \ref{width}b.  This is in accord with what we would expect if it is the rigidity that controls the failure zone width.  Because the triangular lattice remains over-constrained for all finite values of the strut aspect ratio, $x/a$, varying $x/a$ does not change the distance from the threshold of vanishing rigidity and the crack width should remain constant as observed.

There are at least two regimes for all three systems: (i) as $r_i \rightarrow 0$, $w$ reaches the (vertical) system size; (ii) at high $r_i$, $w$ monotonically decreases. Moreover, in both simulation systems we find that when meandering can be excluded, $w \sim r_1^{-1/2}$ in the limit of large $r_1 \equiv \Delta z$.  This is shown in Fig.\ \ref{phenomena}d in all three sample sizes.  In this case, the transverse (horizontal) width of the sample is kept constant so that the extent of meandering cannot increase as (vertical) height is increased.  The data from all three samples sizes coincide in the large $r_1$ limit.  (We also note that in this regime, there is no visual signature of a meandering crack.)  In Fig.\ \ref{width}a, the two regimes are also clearly identifiable for the smallest system where crack meandering is kept to a minimum.  In the experimental system, the behavior is qualitatively similar; however there is too large a spread in the data to make a precise comparison of the functional form.  

To understand this power-law behavior for the intrinsic width of the failure zone, we consider the limit of infinite size where the lateral system boundaries are far from the region of material failure.  In that case, the elastic moduli and the failure of the material of the material are governed by the density of the ``states-of-self-stress'' (or redundant bonds) in the system \cite{Paulrose}.  If there is a large and uniform density of redundant bonds, the system is more rigid than if the states-of-self-stress are sparse.   The most naive approach is to scale two systems with different rigidities so that the density of the states-of-self-stress are the same.  At large $\Delta z$ the density of states-of-self-stress is $\Delta z$. This suggests that in our 2-dimensional systems where the crack is very far from any boundary and the density of bonds is uniform, the area should be normalized by $\Delta z$ so that the linear scale should be normalized by $\Delta z^{1/2}$.  This argument is in accord with the results of the simulations.  

\begin{figure*}
\centering 
\begin{center} 
\includegraphics[width=0.95\columnwidth]{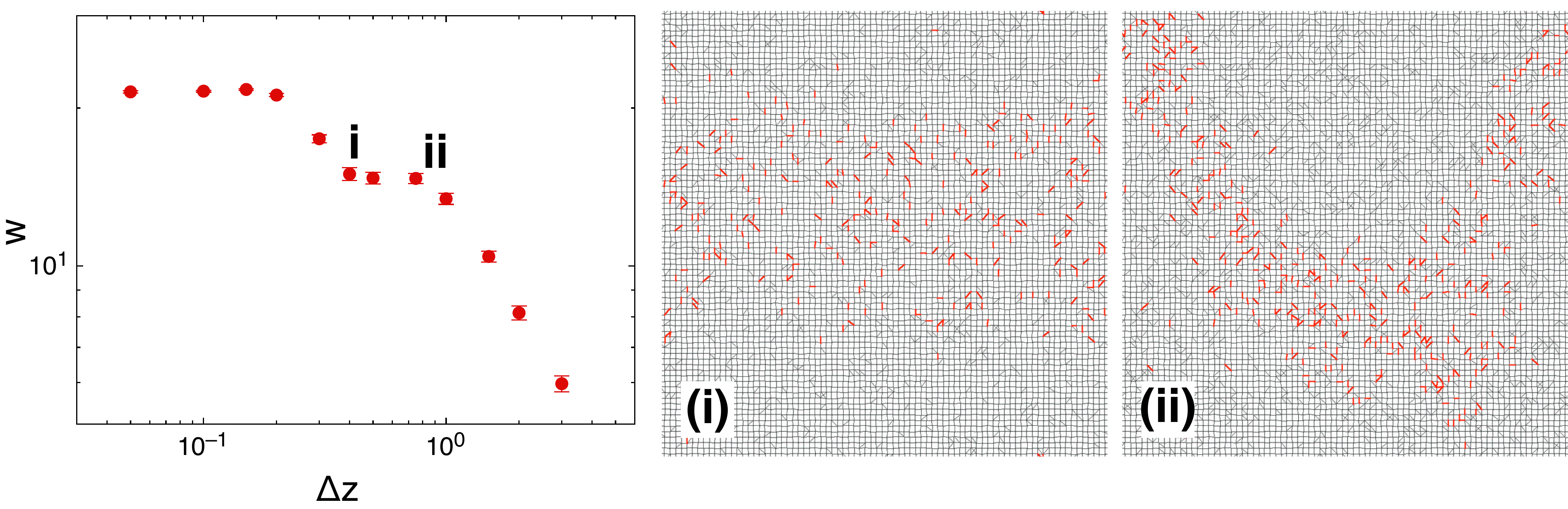} 
\caption{
\label{shoulder}
Crack width vs.\ $\Delta z$ in random-network simulations with a square geometry.  The images are snapshots at the end of the breaking process for two values of $\Delta z$, labeled (i) and (ii) on the plot, illustrating the competition between the intrinsic width and the meandering of a crack.  The intrinsic crack width is initially the size of the sample and then decreases with increasing $\Delta z$.  The image (i) shows when the $w$ has first decreased to a second plateau value.  As $\Delta z$ increases further, the intrinsic width decreases and becomes comparable to the meandering width of the crack.  The image (ii), at the other end of the plateau, shows a meandering crack with a smaller intrinsic width.   At larger $\Delta z$, the entire width of the crack decreases.} 
\end{center}
\end{figure*}

In the intermediate regime between the two asymptotic behaviors, where $w$ first starts to decrease as $r_i$ is increased, the dependence of the crack width on rigidity is more nuanced.  While the asymptotic behaviors at low and high rigidity remain the same, new features emerge in the crossover regime.  This occurs because there are two contributions to our metric for crack width, $w$: the intrinsic width and the width due to meandering of a narrower crack due to disorder. This is demonstrated in Fig.\ \ref{shoulder} where data is shown for the system consisting of a square lattice with cross braces for a larger system size.  Image (i) illustrates a crack with a large intrinsic width.  Image (ii) shows that a narrow crack can also appear wider due to meandering: the crack deviates from a straight path and moves back and forth along the direction transverse to the pulling direction.  In the smallest sample size, the meandering is least important and in the high $\Delta z$ limit one again finds $w \sim r_1^{-1/2}$ (see Fig.\ 2a). 

This cross-over from intrinsic-dominated to meandering-dominated width is most clearly seen in our largest lateral systems, as illustrated in Fig.\ 3.  (In Fig.\ \ref{phenomena}d the transverse (horizontal) size was fixed so that meandering did not play an increased role as system size was varied.)  As described above, our second simulation system (from which the data in Fig.\ \ref{shoulder} is obtained) is based on adding cross-braces to a square lattice.  Thus varying $\Delta z$ controls not only the intrinsic width of a crack by modifying its rigidity, but also changes the amount of disorder.  Upon increasing $\Delta z$ even more, the cross braces are more uniformly distributed throughout the material and the disorder again decreases.  Thus, meandering becomes less important at asymptotically high $\Delta z$.

We have found that similar diffuse failure at low rigidity also occurs in simulations performed under shear as well as under tension.  Due to the small system sizes available in experiments, we are unable to distinguish between meandering and crack-width broadening during material failure.  However, our metric for $w$ (standard deviation in broken bonds) allows us to compare simulations and experiment and show the similar behavior in all the systems we have studied.  Sufficiently close to the rigidity transition, failure naturally occurs in a diffuse manner and will encompass the entire system.  The crossover between  diffuse failure due to rigidity loss and the meandering due to effects of disorder will be explored further in a separate study.

\section{A failure phase diagram}

Shekhawat et al.\ \cite{shek} have previously considered the transition from crack nucleation to damage percolation in a random fuse network with a distribution of burning thresholds $x$ extracted from a probability distribution $F(x)=x^{\beta}$ with $\beta >0$.  In the strong disorder limit, $\beta \rightarrow 0$, damage percolation (e.g., diffusive failure) takes place, while in the weak disorder limit, $\beta \rightarrow \infty$, crack formation occurs.  

\begin{figure}
\centering 
\begin{center} 
\includegraphics[width= 0.7 \columnwidth]{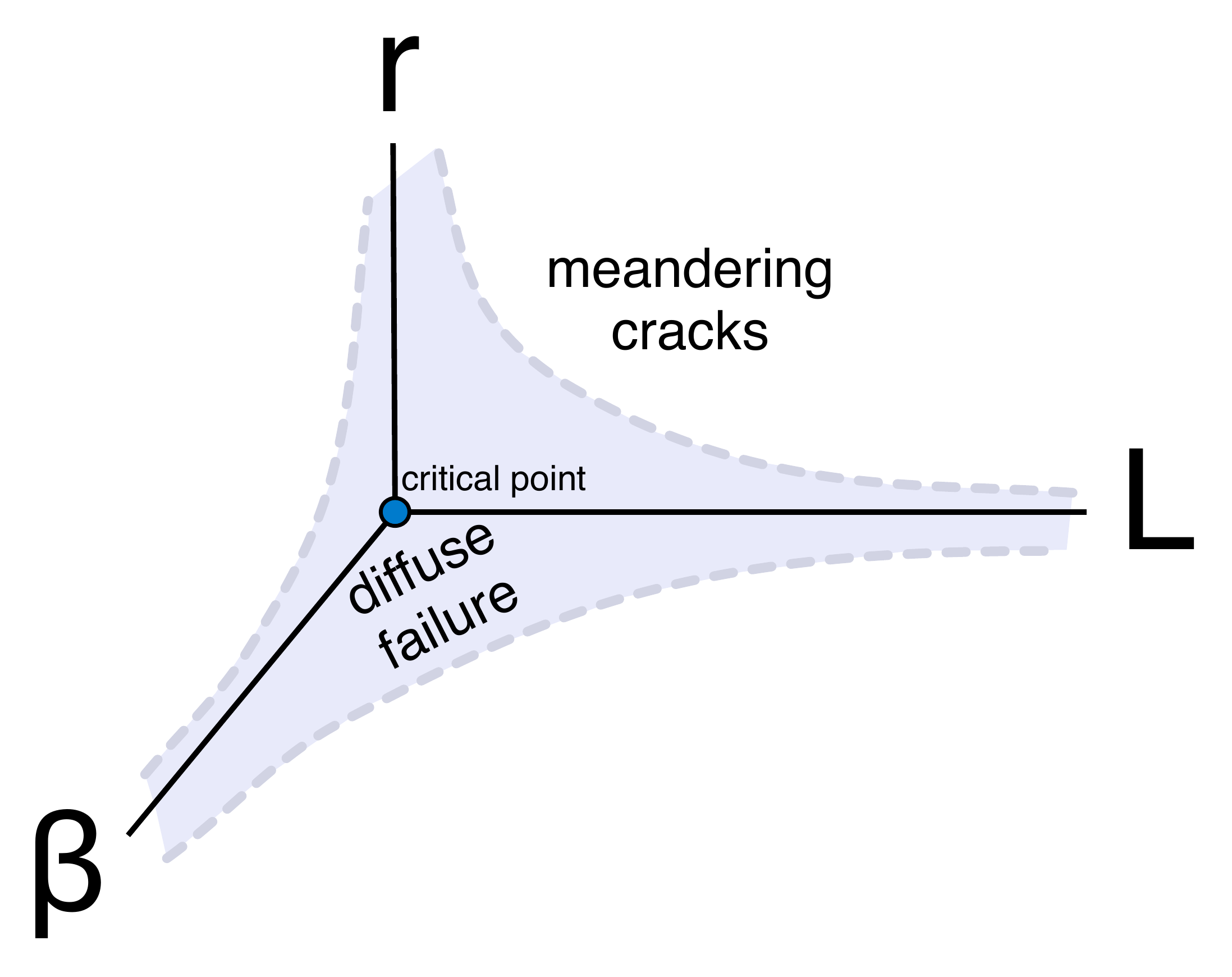}
\caption{
\label{phasediagram}
Schematic fracture phase diagram with three axes: rigidity, $r$; system size, $L$, and inverse disorder, $\beta$.  In addition to $L$ and $\beta$, rigidity controls the transition between diffusive failure and fracture via narrow meandering cracks.}
\end{center}
\end{figure}

In the phase diagram of Fig.\ \ref{phasediagram}, we extend that work to include the effects of material rigidity.  In addition to $\beta$ (inversely proportional to disorder), and $L$ (the system size), we include an axis, $r$, to represent the material rigidity.  As a system becomes less and less rigid, the manner in which it fails changes smoothly from a narrow straight crack, to broad and diffuse failure.  The smooth crossover between regimes is represented schematically where the vertical axis of rigidity  is parameterized by $r$ and the horizontal axis of sample length by $L$.  In the limit $ L \rightarrow \infty$, cracking always wins over diffuse failure except right at the point where rigidity vanishes.

Unlike our models, the random fuse network is agnostic about the rigidity of the underlying lattice, but it does mimic the amount of disorder in the breaking thresholds that we have ignored. This finite-size crossover (indicated by a dashed gray line in the horizontal plane of Fig.\ \ref{phasediagram}) takes place when $L^{\frac{1}{\nu_f}} \beta$ attains a critical value, with $\nu_f \approx 1.5$ \cite{shek}. Figure\ \ref{phasediagram} suggests that the effects of rigidity and disorder can be combined so that different materials can be classified according to their location in a three-dimensional failure ``phase diagram''. 

The key point is that the divergence of the process zone we observe at the threshold of vanishing rigidity $r=0$ mirrors what happens in the infinite disorder limit $\beta =0$. Approaching $r=0$ acts as a magnifying glass for the effect of disorder. This is consistent with the suggestion that the jamming transition is ``the epitome of disorder'', where there is no length scale on which the system can be averaged to regain an elastic description of the solid \cite{ohern}.  Intuitively, the length scale associated with the process zone is a probe of the divergent length scales associated with rigidity loss; failure is the dynamical process that drives the system towards the rigidity transition. Our bond breaking processes unfold along trajectories of decreasing $r$ in this failure phase diagram.  (We note, however, that as bonds break during failure, the distribution of bonds no longer remains uniform.)

In conclusion, as a solid approaches the rigidity transition, its failure behavior changes dramatically.  Although the system ultimately falls apart due to a crack dividing it into two pieces, the nature of this crack is profoundly different at high and low rigidity.  As the system becomes less and less rigid, the crack becomes wider and wider until the width of the crack reaches the system size.  In this regime, the bonds initially break at apparently random positions until they produce a percolating cluster of broken bonds across the sample. Because the spatial extent of the failure process zone depends on material toughness, varying the rigidity can be used as a lens to examine the non-linear response that would otherwise be observable only on a microscopic scale in a rigid material.

\bigskip

We acknowledge inspiring interactions with L.\ Mahadevan and Nitin Upadhyaya in the
initial stages of this project. We thank Carl Goodrich, Andrea Liu, Michael Marder and Jim Sethna for many fruitful discussions. The work of MMD and SRN was supported by the NSF grant DMR-1404841. Use of facilities of the University of Chicago NSF-MRSEC are gratefully acknowledged. BGC, THB and SU thank FOM and NWO for financial support.


\end{document}